# The Critical Metallization of Hydrogen in Pressurized LaBeH$_8$ Hydride

Zihan Zhang[1], Tian Cui[2, 1, *], Yansun Yao[3], Tianchen Ma[1], Qiwen Jiang[1], Haojun Jia[4], Bartomeu Monserrat[5, 6], Chris J. Pickard[6, 7], Defang Duan[1, *]

[1]*State Key Laboratory of Superhard Materials and Key Laboratory of Material Simulation Methods & Software of Ministry of Education, College of Physics, Jilin University, Changchun 130012, China*

[2]*Institute of High Pressure Physics, School of Physical Science and Technology, Ningbo University, Ningbo, 315211, People's Republic of China*

[3]*Department of Physics and Engineering Physics, University of Saskatchewan, Saskatoon, Saskatchewan, Canada, S7N 5E2*

[4]*Department of Chemistry, Massachusetts Institute of Technology, Cambridge, MA, 02139, United States.*

[5]*Department of Materials Science & Metallurgy, University of Cambridge, 27 Charles Babbage Road, Cambridge CB3 0FS, United Kingdom*

[6]*Cavendish Laboratory, University of Cambridge, J. J. Thomson Avenue, Cambridge CB3 0HE, United Kingdom*

[7]*Advanced Institute for Materials Research, Tohoku University 2-1-1 Katahira, Aoba, Sendai, 980-8577, Japan*

**Abstract:**

Behaviours of hydrogen, such as fluidity and metallicity, are crucial for our understanding of planetary interiors and the emerging field of high-temperature superconducting hydrides. These behaviours were discovered in complex phase diagrams of hydrogen and hydrides, however, the transition mechanism of behaviours driven by temperature, pressure and chemical compression remain unclear, particularly in the processes of metallization. Until now, a comprehensive theoretical framework to quantify atomization and metallization of hydrogen in phase diagram of hydrides has been lacking. In this study, we address this gap by combining molecular dynamics and electronic structure analysis to propose a theoretical framework, which clarify the content and properties of atomic hydrogen under various temperature and pressure conditions and chemical compression exerted by non-hydrogen elements in hydrides. Applying this framework to the superconducting hydride $LaBeH_8$, we identify three general hydrogen orderings within its phase diagram: molecular, sublattice and warm hydrogens. During the phase transition from molecule to sublattice, hydrogen exhibits different properties from three general hydrogen orderings, such as fast superionicity, metallicity and unusual atomic content response to temperature. These abnormal behaviours were defined as the critical metallization of hydrogen, which not only suggests a potential synthesis route for the metastable phase but also provides valuable insights into the complex synthetic products of superconducting hydrides.

**Introduction**

Hydrogen, the most abundant element in the universe and the lightest in mass constitutes approximately 75% of all visible matter. Hydrogen generally exists in complex environments under extreme conditions[1], and the resulting hydrides exhibit properties such as superionicity[2-4] and high-temperature superconductivity[5,6].

The importance of the chemical environment of hydrogen can be illustrated with the example of hydrogen-bond symmetrization driven by the large kinetic energy of hydrogen, has been observed during the ice phase VII-to-X transition at around 80 GPa[7]. In this context, introducing small amounts of salt into ice can significantly modify the properties of hydrogen-bond symmetrization, while cations $Li^+$ or $K^+$ can increase the pressure required for hydrogen-bond symmetrization by 30 GPa[8,9], cation $Ca^{2+}$ conversely decreases the pressure by 30 GPa, and exhibits "chemical pressure"[10]. As another illustration, chemical compression in hydrides offers the potential to achieve high-pressure phenomena of hydrogen at lower pressures. Two notable high-pressure phenomena of hydrogen are fluidity and superconductivity[1]. Under chemical compression, the diffusion constant of superionic states in ice reaches a maximum in the boundary of hydrogen-bond symmetrization in the VII-to-X phase transition[11,12]. Additionally, chemically confined water (such as within a nanotube) can enter superionic states near ambient conditions[13,14]. The concept of chemical compression has motivated the pursuit of high-temperature superconductivity in hydrogen-rich compounds through metallized hydrogen species in metal lattices [5,6,15-20]. However, unlike pressure and temperature, which have clear thermodynamic definitions, the effects of complex chemical environments on hydrogen are poorly understood, and a general phase diagram of hydrogen under chemical compression remains unclear.

The transition of insulating molecular hydrogen into a metallic state is believed to be achievable under extreme $P$-$T$ conditions. Still, the required conditions would exceed 500 GPa[21,22] and thousands of Kelvin[23-28], posing extreme experimental difficulties. Alternatively, chemical compression can significantly reduce the pressure needed for the molecule-to-atomic transition of hydrogen in hydrogen-rich compounds[5,6,15-20], an observation that triggered the discovery of high-temperature superconducting hydrides. For example, $Mg_2IrH_6$ has recently been proposed as a superconducting hydride at ambient pressure with a critical temperature ($T_c$) up to 160 K[29,30]. Superconducting hydrides are ideal candidates to clarify the effects of complex chemical environments on the properties of hydrogen. It is possible to identify qualitatively consistent one-to-one mappings of properties between pure hydrogen and chemically compressed hydrogen; for instance, the melting line of pure hydrogen[31,32] corresponds to the superionic lines of hydrides[2-4,33,34], and the atomic phase of hydrogen[35,36] corresponds to the atomic hydrogen sublattice in hydrides[5,6,15-20]. However, the phase diagram analysis for chemically compressed hydrogen is much more complex than that of pure hydrogen, and understanding of chemically compressed hydrogen lags behind that of pure hydrogen. For example, recent research has indicated the supercritical behaviour in liquid hydrogen during the insulator-metal transition at high pressure[37], but the potential analogous critical behaviours of chemically compressed hydrogen during metallization remain unclear. Moreover, the superionic states of hydrogen in hydrides exhibit more complex diffusion phenomena than pure hydrogen due to the rearrangement of molecular orbitals via geometrical confinement and electron transfer from the "compressors".

In this context, it would be desirable to explore dynamic phenomena of hydrogen under these complex conditions. The recently discovered $LaBeH_8$ hydride[38,39] provides a promising platform for dynamic and electronic research of superconducting hydrides [40], as it exhibits a stable phonon dispersion at 20 GPa, significantly below its thermodynamic stable pressure of 98 GPa ($P_{ts}$). Therefore, the La-Be lattice offers an extensive pressure range for studying hydrogen behaviour under chemical compression, making $LaBeH_8$ an ideal candidate for this research.

In this study, we investigate the interplay between three pathways to hydrogen metallization: pressure, temperature, and chemical compression, as illustrated in FIG 1. The underlying mechanism for hydrogen metallization through pressure and chemical compression is similar. When subjected to high pressure, wavefunctions overlap between adjacent hydrogen molecules increases, enhancing intermolecular interaction that leads to molecular dissociation and triggers the transition to a metallic state. Similarly, chemical compression enhances wavefunction overlap between hydrogen and non-hydrogen elements, causing the breakdown of molecular hydrogen and resulting in the metallization[41]. In contrast, temperature-driven metallization is initiated by the kinetic energy imparted to hydrogen molecules, which first dissociates and subsequently transit into a warm liquid phase[28]. To elucidate the interactions among these three pathways to hydrogen metallization, we analyze the microscopic dynamical behaviours of hydrogen and apply the findings to the phase diagram of the representative hydride LaBeH$_8$.

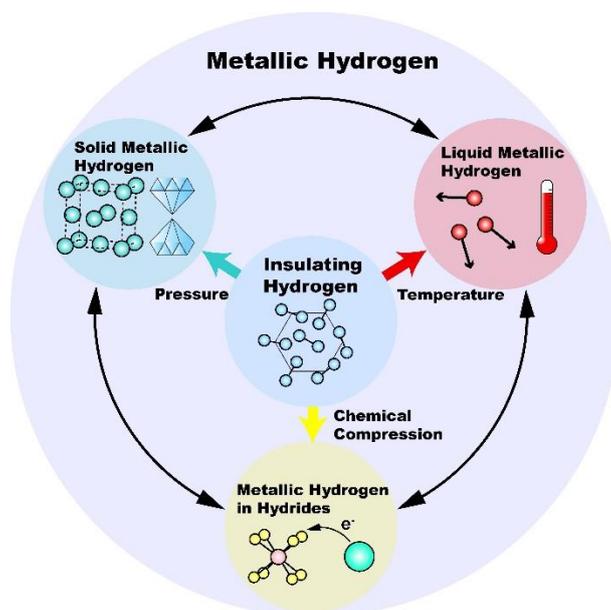

FIG 1| Schematic representation of the three distinct routes to hydrogen metallization from its insulating state. The blue, red, green, and yellow regions denote insulating hydrogen, temperature-driven metallization, pressure-driven metallization, and chemical-compression-driven metallization, respectively. The blue, red, green, and yellow spheres denote molecular hydrogen, liquid atomic hydrogen, solid atomic hydrogen, and atomic hydrogen within a hydride sublattice, respectively. The large green and pink spheres in the yellow region correspond to La and Be elements in LaBeH$_8$, respectively.

**Molecular Dynamics Simulation of Hydrogen Orderings in Hydrides**

We carried out molecular dynamics simulations to clarify the hydrogen ordering in molecular, sublattice and warm states within LaBeH$_8$ hydride at various pressure-temperature ($P$-$T$) points as depicted in FIG 1. The radial distribution functions (RDFs) of LaBeH$_8$, derived from molecular dynamics trajectories at three representative $P$-$T$ points, each reflecting a distinct route to hydrogen metallization, are shown in FIG. 2.

In FIG 2a, the first route involves increasing pressure (represented by increasing density) at a constant temperature of 600 K. When the pressure exceeds a characteristic threshold of $P_{ts}$ (~ 98 GPa), LaBeH$_8$ hydride becomes thermodynamically stable relative to its constituent products. Below thermodynamically stable pressure $P_{ts}$, most hydrogen in the system exists in its molecular form. As pressure increases, the proportion of hydrogen with short H-H bonds (~ 0.8 Å) decreases, while the proportion with longer bonds (~ 1.5 Å) increases. This transition suggests the dissociation of molecule hydrogen, leading to the formation of an atomic sublattice within the hydride and reaching a metallic state. The second and third routes involve increasing temperature at two constant pressures, one below and one above $P_{ts}$, as illustrated in FIGs 2b and 2c. Below $P_{ts}$, at a density of 4.97 g/cm$^3$, LaBeH$_8$ tends to decompose into molecular hydrogen, metals, and lower hydrides. In this case, the quantity of hydrogen molecules (indicated by short H-H bonds) decreases as temperature rises (FIG 2b), indicating the thermal breakdown of hydrogen molecules (FIG. S6a-S6c) to reach a metallic state. However, high temperatures do not facilitate the formation of LaBeH$_8$ hydride; instead, they cause distortions within the system. In contrast, above $P_{ts}$, at a density of 7.20 g/cm$^3$, LaBeH$_8$ is thermodynamically stable, but high temperatures can lead to the decomposition of the hydride. This is evidenced by the increased presence of hydrogen with short H-H bonds at elevated temperatures (FIG 2c). Detailed snapshots of these scenarios are provided in FIGs. S6a-S6o.

We performed a Bader charge analysis (BCA) on snapshots from molecular dynamics trajectories (FIGs. S22-S26) to further understand these observations. In the molecular state, *i.e.*, when the system is pressurized below $P_{ts}$, a non-equilibrium process of "chemical pre-compression" is observed: electrons from La and Be transfer to nearby molecular hydrogens, populating the antibonding orbital and lengthening the H-H bond. In the sublattice state, *i.e.*, when LaBeH$_8$ is thermodynamically stable, nearly all hydrogens exhibit charge values in the range 1.2-1.4 e$^-$, which indicates a negatively charged hydrogen sublattice. In the warm hydrogen state, the maximum and minimum values of charge showed a linear correlation of H-H bond length, regardless of pressure (FIGs. S22-S26). Therefore, warm hydrogen in hydride LaBeH$_8$ at different pressure exhibits similar charge distribution, which was constructed by relatively uniform arrangements of atomic hydrogen, ionic hydrogen and hydrogen clusters such as H$_2^+$ and H$_3^+$ cations [42] in hydrogen plasma. Thus, temperature disrupts hydrogen ordering in both molecular and hydride states, pushing hydrogen towards a warm hydrogen state. A similar temperature-dependent behaviour of hydrogen ordering was also observed in hydrides LaH$_{10}$[43] and NH$_3$[44], suggesting a potential general behavior for hydrogen-rich compounds.

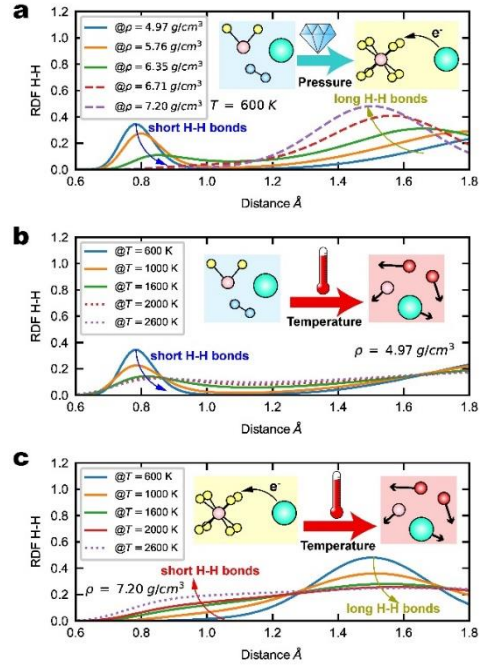

FIG 2| Radial distribution function (RDF) illustrating the changes in hydrogen ordering of LaBeH$_8$ hydride under different stimuli from machine learning molecular dynamics. (**a**) Evolution of RDF under increasing pressure (density) at approximately 600 K. Solid and dashed lines represent metastable and thermostable initial structures under static pressure, respectively. (**b**) Evolution of RDF under increasing temperature at constant density ~ 4.97 g/cm$^3$, where the initial structure is metastable. Solid and dotted lines indicate solid and liquid phases, respectively. (**c**) Evolution of RDF under increasing temperature at constant density ~ 7.20 g/cm$^3$, where the initial structure is thermodynamically stable. Solid and dotted lines denote solid and liquid phases, respectively. Insets: Blue, red, and yellow represent hydrogen in insulating, temperature- and chemical-compression-driven metallic states, respectively. Blue, red, and yellow spheres represent molecular hydrogen, atomic hydrogen liquid, and atomic hydrogen in the sublattice of hydrides, respectively. The large green and pink spheres denote La and Be elements, respectively.

**Quantification hydrogen dissociation**

The H-H bond length is the key quantity for describing the dissociation of hydrogen molecules and the associated insulator-to-metal transition of hydrogen in the hydrides[37]. We introduce a method to quantify hydrogen dissociation in hydrogen-rich compounds by analyzing the H-H bond length within the hydrogen sublattices.

Chemical compression can induce the formation of hydrides with a higher hydrogen content than expected based on the valency of the non-hydrogen elements. For instance, lanthanum typically has a maximum valence of +3; therefore, the expected lanthanum hydride is $LaH_3$. Thus, in $LaH_{10}$, three hydrogen atoms are 'expected' based on valency, while the remaining seven are considered 'additional'. These additional hydrogen atoms are 'chemically compressed' by the La sublattice, reaching a metallic state in $LaH_{10}$ at significantly lower pressures than required for pure hydrogen to metalize. This rationale allows one to describe $LaH_{10}$ as $(LaH_3)H_7$, as suggested by recent simulations[43]. Similarly, $LaBeH_8$ may be described as $(LaBeH_5)H_3$. Hydrogen atoms are usually arranged in a high-symmetry sublattice in hydrides. When the structure is distorted, the ability of chemical compression to stabilize atomic hydrogen diminishes, causing hydrogen to form molecular structures with short H-H bonds.

We define $I_{H_2}$ as the volume integral of the atomic distribution $g(r)$ within a characteristic H-H bond length $r_{H-H}$,

$$I_{H_2} = 4\pi \int_0^{r_{H-H}} g(r)\, r^2 dr. \qquad (1)$$

$I_{H_2}$ represents the average coordination number of hydrogen atoms within $r_{H-H}$, equal to 1 for molecular hydrogen and 0 for atomic or ionic hydrogen, where hydrogen atoms are further separated. When all 'additional' hydrogens form molecules and 'expected' hydrogens are ionic, $I_{H_2} = \frac{N_{aH}}{N_H}$, where $N_{aH}$ and $N_H$ denote the number of additional and total hydrogen atoms, respectively. The relative molecular rate is defined as $R_{H_2} = I_{H_2} \times \frac{N_H}{N_{aH}}$. If $R_{H_2} = 0$, all hydrogen atoms are in atomic form. If $R_{H_2} = 1$, 'additional' hydrogens form molecules and 'expected' hydrogens become anions. If $R_{H_2} > 1$, the 'expected' hydrogens form molecules. We refer to $1 - R_{H_2}$ as the relative atomic rate (RAR). This framework is element-independent and applicable to any hydride with 'additional' hydrogens. The relative atomic rate can be used in molecular dynamics simulations to reflect the time-averaged molecular dissociation of 'additional' hydrogens, providing insights into the dynamic nature of chemical compression in hydrides. To investigate the relative atomic rate $1 - R_{H_2}$ for chemical compression using eq. (1), we must determine $r_{H-H}$, the onset length for molecular dissociation. It is logical that $r_{H-H}$ depends on the bond strength change under pressure. To this end, we employed the electron localization function (ELF) as a measure of bond strength to determine $r_{H-H}$. In FIGs S12-16, the ELF values at the midpoint between two hydrogen atoms from dynamic simulation snapshots are plotted as functions of the H-H distances. In most cases, the ELF slope changes abruptly from a gradual decrease around 1.0 to a steep descent toward 0. The point where the ELF slope changes is identified as $r_{H-H}$, indicating a transition from bonding to dissociation. As density decreases from 4.97 $g/cm^3$ to 7.20 $g/cm^3$, the onset length for molecular dissociation also decreases from 0.95 Å to 0.85 Å. We use these varying $r_{H-H}$ values at different densities for calculations of $R_{H_2}$.

**Critical metallized hydrogen**

We calculate the RAR under various *P-T* conditions at five different densities to investigate the dissociation of hydrogen during metallization of LaBeH$_8$ hydride, as shown in FIG 3. The results show distinct behaviours of RAR below and above $P_{ts}$. Above $P_{ts}$, RAR primarily indicates atomic ordering within the sublattice with a maximum value close to 99.9%. Below $P_{ts}$, RAR suggests molecular ordering, with a minimum value of around 7.7%. Additionally, RAR exhibits a local maximum in the region around $P_{ts}$ below 1800 K, which is the boundary of molecular to atomic phase and suggestive of critical metallization. Critical phenomena typically emerge during phase transitions, such as the liquid-gas transition of water, molecule-to-atomic transition of liquid hydrogen[37], and (anti)ferromagnet-paramagnet transition of correlated electrons, where multiple phases coexist, and boundaries vanish. Alongside changes in macroscopic quantities like heat capacity and density, the electronic structure provides a microscopic probe for these critical behaviours[45,46]. In our model, we used electron localization, or its absence, to measure hydrogen metallization, which is reflected in RAR. In this work, the RAR for critically metallized hydrogen exhibits an unusual response to temperature (FIG 3). For the pressures close to $P_{ts}$ (98 GPa[38] for LaBeH$_8$), RAR initially increases to a maximum value (57.9% at 1800 K and 82.3% at 800 K for densities of 5.76 and 6.35 $g/cm^3$, respectively) and then decreases as temperature rises. This behavior contrasts sharply with the monotonic response of RAR to temperature observed in molecular and sublattice orderings.

A typical critical behaviour of liquid-gas transition is the formation of a supercritical fluid. Therefore, we investigated the diffusion coefficients $D_H$ associated with the superionic behaviour of hydrogen[47] in its molecular, sublattice and critical metallized states. We fitted the diffusion barrier of superionicity below the melting line with different densities as shown in FIG S45-S49, and found that energy barriers of superionicity are 0.48, 0.40, 0.25, 0.33, 0.51 eV at densities 4.97, 5.76, 6.35, 6.71 and 7.20 $g/cm^3$, respectively. Notably, critical metallization of hydrogen reduce the barriers of superionicity to minimum at density 6.35 $g/cm^3$ (close to $P_{ts}$ = 98 GPa as shown in FIG 3). We use a threshold value of $D_H = 10^{-5}\ cm^2/s$ to distinguish temperature-driven superionicity, as quantum effects typically result in $D_H$ around $10^{-6}$ $cm^2/s$[33,34], and therefore the $D_H$ exceeding $10^{-5}$ $cm^2/s$ indicates a thermal origin for superionicity. In the critical region, hydrogen exhibits the highest $D_H$ compared to other pressures at the same temperature, suggesting a superionic state driven by critical metallized hydrogen. Hydrogen metallization via molecular dissociation reduces intra-molecular interactions while enhancing inter-molecular interactions, thereby facilitating hydrogen transport. Moreover, the observed decrease in the melting point with increasing pressure is consistent with behavior seen in alkali metals such as lithium[48,49], sodium[50] and potassium[51,52]. Our findings indicate that chemically compressed hydrogen in the critical regime exhibits properties akin to atomic hydrogen, with valence electrons resembling those of alkali metals. As pressure further increases beyond the critical region, the hydrogen sublattice is stabilized thermodynamically, with an RAR approaching unity, leading to an increase in the superionic transition temperature.

We further compared the phase diagram of critically metallized hydrogen with critical behaviours observed in water, metallized pure hydrogen, and correlated electron systems. The hydrogen phase diagram within the LaBeH$_8$ hydride (FIG 3 **b**) shows greater similarity to the phase diagram of correlated electron systems, such as heavy fermion superconductors[53] (FIG 3 **c**). The energy scale of quantum nuclei (~ 100 meV) in hydride superconductors is significantly larger than that of strongly correlated electrons (~ 10 meV) in heavy fermion superconductors[54,55]. Consequently, hydride superconductors' pressure and temperature scales (100 GPa and 100 K) are much larger than those for heavy fermion superconductors (1 GPa and 1 K). At low pressures, hydrogen and correlated electrons exhibit localized behavior within their respective scales, with hydrogen displaying molecular ordering and electrons displaying magnetic order. As pressure increases, hydrogen and correlated electrons become nonlocal, transitioning to a sublattice ordering and a Fermi liquid phase. All ordered phases can be disrupted by increasing temperature, and a critical point emerges in the transition between two ordered phases in both phase diagrams.

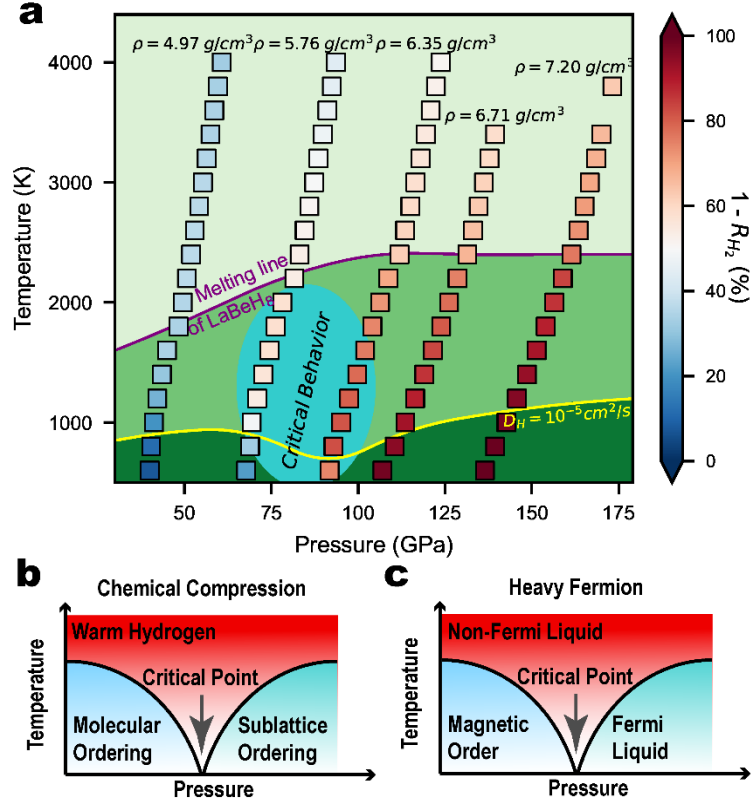

FIG 3| **a** Phase diagram for the LaBeH$_8$ hydride showing pressure (density), temperature and relative atomic rate $(1 - R_{H_2})$. The color of the squares represents the relative atomic rate derived from molecular dynamics simulations. The yellow and purple curves indicate the lines corresponding to diffusion coefficients $D_H = 10^{-5}\ cm^2/s$, and the melting line, respectively. **b** and **c** Schematics of phase diagrams of hydrogen under complex chemical environments and for correlated electrons.

**Metallic hydrogen in the phase diagram of LaBeH$_8$**

To elucidate the connection between metallization and dynamic behaviour of hydrogen in the phase diagram of LaBeH$_8$ hydride (FIG 3), we calculate the vibrational density of states (VDOS) and projected density of states on hydrogen (PDOS on H) from AIMD simulations. VDOS peaks at high frequencies (>100 THz), corresponding to H-H bond stretching, are indicative of molecular hydrogen. These peaks are observed in the VDOS of cold molecular hydrogen at low pressure (density ~ 4.97 g/cm$^3$) and 500 K, as shown in FIG 4a. At a pressure where LaBeH$_8$ is dynamically stable (density ~ 7.20 g/cm$^3$), VDOS peaks appear around 58 THz, indicating the formation of a hydrogen sublattice. As temperature increases, these VDOS peaks smear out, suggesting a temperature-driven transition from molecular and sublattice hydrogen to warm hydrogen, characterized by the absence of high-frequency molecular peaks at any pressure. Next, we investigated the metallization of hydrogen using the PDOS of H (FIG 4**b**, FIG S7~S11). An increase in PDOS with increasing temperature signifies the metallization of hydrogen in LaBeH$_8$, as shown in FIG 4**b**. Notably, the metallization of hydrogen in metastable LaBeH$_8$ coincides with the disappearance of the VDOS peaks of molecular hydrogen. For thermostable LaBeH$_8$, although warm hydrogen exhibits shorter H-H bonds than hydrogen at low temperatures, the PDOS of H at the Fermi level remains almost unchanged with increasing temperature. This suggests that the temperature-driven emergence of short H-H bonds does not weaken the metallization. Short H-H bonds in both molecular hydrogen and warm hydrogen exhibit similar ELF values at the bond centers (FIG S12~S16) but differ in their metallization properties. Thus, high temperature does not alter the wavefunction overlap mechanism of chemical bonding but broadens the energy level for metallization. This leads to a wider broadening of ICOHP at high temperatures compared to low temperatures, as shown in FIG S17~S21. In the critically metalized region, the PDOS of hydrogen is higher than that of molecular ordering but lower than sublattice ordering with VDOS peaks of molecular hydrogen, exhibiting intermediate metallization of hydrogen in hydride LaBeH$_8$.

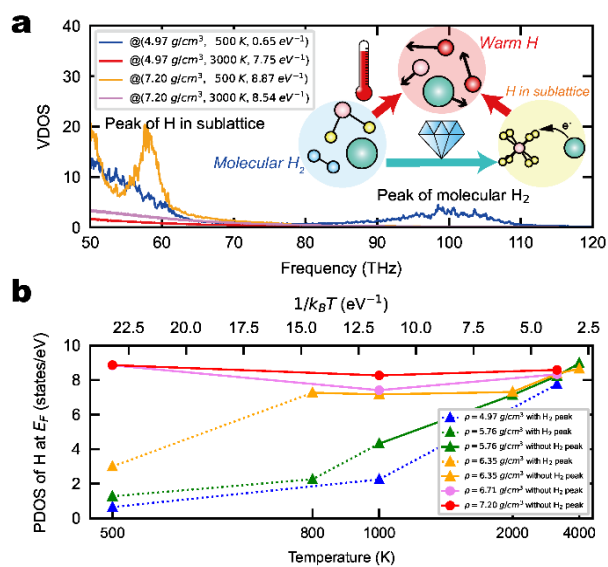

FIG 4| Vibrational density of states (VDOS) of hydrogen and projected density of electronic states (PDOS) on hydrogen at Fermi level $E_F$ of random snapshots from AIMD trajectories. **a** VDOS of hydrogen at various densities and temperatures. The legend indicates density, temperature, PDOS on hydrogen at Fermi level $E_F$. **b** Average PDOS of hydrogen at Fermi level $E_F$ from random snapshots as a function of temperature. Solid and dotted lines

represent VDOS without and with $H_2$ peaks, respectively. Triangles and circles indicate densities (pressures) corresponding to the metastability and thermodynamic stability of hydride $LaBeH_8$, respectively.

**Summary**

In summary, we investigated the behaviours of hydrogen under chemical compression in $LaBeH_8$ hydride. By combining analyses of electronic structure and dynamic properties, our approach quantified the degree of hydrogen metallization in complex chemical environments, identifying four distinct regions: molecular hydrogen, sublattice hydrogen, warm hydrogen, and critical metalized hydrogen. The properties of these regions are detailed in Table 1. The critical metalized behaviour of hydrogen suggests a local maximum of relative atomic rate, driven by the interplay of temperature and chemical compression below $P_{ts}$. Our results reveal an analogy between the critical behaviours of chemically compressed hydrogen and correlated electrons.

Table 1| Summary of behaviours of molecular, sublattice, warm, and critical metalized hydrogens.

| Behaviors of hydrogen | Molecular hydrogen | Sublattice hydrogen | Warm hydrogen | Critical metalized hydrogen |
|---|---|---|---|---|
| Metallicity from PDOS of H | Bad metallicity | Good metallicity | Good metallicity | Medium metallicity |
| VDOS peaks of hydrogen | High frequency (> 100 THz) | Low frequency (< 80 THz) | No peaks of vibration | Weak peaks at high frequency |
| Diffusion coefficients of H | Low | Low | Liquid | Faster than those in molecular and sublattice hydrogen at the same temperature |
| Minimum length of H-H bonds | ~ 0.8 Å | ~ 1.2 Å | ~ 0.8 Å | ~ 0.8 Å |
| Pressure and temperature conditions | Below $P_{ts}$, low temperature | Above $P_{ts}$, low temperature | High temperature | Around $P_{ts}$, low temperature |

The conditions of laser hearting for synthesizing hydrides near the critically metalized hydrogen region indicate the presence of critical behaviour—where molecular hydrogen decomposes to atomic hydrogen, resulting in a significant increase in the diffusion coefficient. This facilitates the synthesis of metastable phases. For instance, complex metastable phases of $LaH_x$[56-58] and $YH_x$[59] can be synthesized with metal elements and $NH_3BH_3$ after laser heating at pressures lower than those of $LaH_{10}$ and $YH_6$. Furthermore, a comprehensive description of the complex dynamic behaviour of hydrogen, governed by temperature, pressure and chemical pressure, not only provides new insights and guidance for future experiments on superconducting hydrides but also highlights the rich properties of hydrogen for designing novel materials, such as fast hydrogen conductors. The behaviours of critically metallized hydrogen in LaBeH$_8$ hydride could be confirmed using recently developed high-pressure nuclear magnetic resonance spectroscopy[60,61]. Moreover, a fast proton transport phase with maximal diffusion coefficient was also observed during the VII-to-X transition of ice, where hydrogen bonds are symmetrized [11,12]. This suggests that critical behaviours of hydrogen might be overlooked in various phase transition phenomena in complex chemical environments.

**Acknowledgements**

This work was supported by National Key R&D Program of China (No. 2022YFA1402304), National Natural Science Foundation of China (Grants No. 12274169, 12122405 and 52072188), the Program for Science and Technology Innovation Team in Zhejiang (No. 2021R01004) and the Fundamental Research Funds for the Central Universities. B.M acknowledges support from a UKRI Future Leaders Fellowship [MR/V023926/1], from the Gianna Angelopoulos Programme for Science, Technology, and Innovation, and from the Winton Programme for the Physics of Sustainability. C.J.P. acknowledges financial support from the Engineering and Physical Sciences Research Council [Grant EP/P022596/1] and a Royal Society Wolfson Research Merit award. Parts of the calculations were performed in the High Performance Computing Center (HPCC) of Jilin University and TianHe-1(A) at the National Supercomputer Center in Tianjin.